\documentclass[twocolumn,english]{revtex4-2}
\usepackage[T1]{fontenc}
\usepackage[utf8]{inputenc}
\usepackage{color}
\usepackage{babel}
\usepackage{bm}
\usepackage{amsmath}
\usepackage{amssymb}
\usepackage{graphicx}
\usepackage[unicode=true,pdfusetitle,
 bookmarks=true,bookmarksnumbered=false,bookmarksopen=false,
 breaklinks=false,pdfborder={0 0 1},backref=false,colorlinks=true]
 {hyperref}

\makeatletter
\renewcommand\[{\begin{equation}}
\renewcommand\]{\end{equation}}
\usepackage{xcolor}
\hypersetup{
     colorlinks   = true,
     citecolor    = blue,
     urlcolor     = blue,
     urlbordercolor = blue
}

\makeatother

\usepackage[a4paper]{geometry} 
\begin{document}
\title{Proposal of quantum arrival-time measurement with a Bose-Einstein
condensate}
\author{Pascal Naidon}
\email{pascal@riken.jp}

\author{Lucas Happ}
\altaffiliation{Present address: Institut für Quantenphysik and Center for Integrated Quantum Science and Technology (IQST),Universität Ulm, D-89069 Ulm, Germany}

\affiliation{Few-Body Systems Physics Laboratory, RIKEN Nishina Centre, RIKEN,
Wakō, 351-0198 Japan}
\author{Denis Boiron}
\affiliation{Université Paris-Saclay, Institut d’Optique Graduate School, CNRS,
Laboratoire Charles Fabry, 91127, Palaiseau, France}
\date{\today}
\begin{abstract}
This work shows how a Bose-Einstein condensate of ultracold atoms
could be used to address a long-standing question in quantum theory:
how much time does it take for a particle to reach a detector? To
this end, we propose a realistic experimental setup, whose key idea
is not to measure arrival times directly, but the arrival flux on
the detector as a function of its position. This novel approach not
only solves practical issues with having a detector close to the system,
but also results in signals that allow to unambiguously distinguish
different theoretical predictions. This proposal raises prospects
for resolving the decades-old debate on this fundamental issue. 

\end{abstract}
\maketitle
Ultracold atoms have been an essential platform to verify numerous
predictions of quantum theory, Bose-Einstein condensation~\citep{Pethick2008,Castin2025}
being the most prominent and famous example. All observations done
so far have constituted confirmations of the quantum theory's predictions,
such as the spectacular observation of matter wave interference~\citep{Andrews1977}.
There is however a seemingly simple question for which quantum theory
has long been known~\citep{Muga2000} for not giving a clear answer:
how much time does it take for a particle to go from one point to
another? Due to the statistical nature of quantum theory, the question
should of course be understood in terms of the distribution of times
of arrival of the particle. This question has a long history~\citep{Pauli1958,Aharonov1961,Paul1962,Allcock1969a,Allcock1969b,Allcock1969c,Kijowski1974,Werner1987,Grot1996,Delgado1997,Aharonov1998,Leavens1998,Muga1999,Aoki2000,Baute2000,Wlodarz2002,Galapon2005,Anastopoulos2006,Nitta2008,Anastopoulos2012,Giovannetti2015,Maccone2020,Tumulka2022a,Juric2022,AyatollahRafsanjani2023,Roncallo2023,Anastopoulos2023,TahvildarZadeh2024,Araujo2024,Naidon2024,Beau2024,Beau2025,Reddiger2026}
marked by controversies~\citep{Kijowski1999,Leavens2002,Egusquiza2003,Leavens2005,Das2019,Das2021,Goldstein2024,Das2023,Goldstein2024b,Cavendish2024,Maccone2025}.
Authors even disagree about whether the standard formulation of quantum
theory provides an answer to the question or requires to be extended.
At the very least, the answer does not appear to be straightforward,
and several proposals have been made, leading to distinct predictions
of arrival-time distributions. To our knowledge, no experiment has
been conducted yet to explicitly test these predictions. Experiments
done so far~\citep{Robert2001,Narevicius2008,Dall2011,Dall2013,Stopp2021}
that have measured arrival times of a particle were performed in a
regime where all predictions agree. The problem has thus remained
mostly a theoretical issue, although many simple experiments have
been proposed by theoreticians~\citep{Muga2000a,Das2019,Roncallo2023,AyatollahRafsanjani2023,Naidon2024}.
While these proposals are feasible in principle, none of them could
be realized so far, due to various practical limitations, most notably
the necessity to detect the particle very close to its source. There
is thus a strong need to propose a readily achievable experiment that
can finally settle the issue empirically.

One of the prominent candidates for such arrival-time experiments
in the quantum regime are ultracold atoms, due to their versatility
and controllability~\citep{Pethick2008,Castin2025}. In particular,
atomic Bose-Einstein condensates present the advantage of having many
identical atoms in the same quantum state, thus providing an efficient
way to perform at once many arrival-time measurements on a single
quantum state. This letter builds upon this idea to propose a realistic
experimental setup whereby different theoretical predictions of arrival
times could be unambiguously discriminated.

\paragraph*{Overview of arrival-time predictions ---}

We consider four predictions of the arrival-time distribution of a
non-relativistic quantum particle of mass $m$, derived from four
theoretical approaches:
\begin{enumerate}
\item \emph{The quantum clock approach}: this approach~\citep{Giovannetti2015,Maccone2020,Roncallo2023}
originates from the idea~\citep{Page1983,Aharonov1984} that time
itself should be treated quantum-mechanically, as the observed position
of a clock hand that is ruled by the laws of quantum theory. According
to this approach, the distribution $\Pi(t)$ of times $t$ for the
particle to arrive onto a surface $S$ is simply proportional to its
probability of presence on the surface, i.e.
\begin{equation}
\Pi(t)=\alpha\iint_{S}\vert\psi(\bm{r},t)\vert^{2}dS,\label{eq:density}
\end{equation}
where $\psi(\bm{r},t)$ is the wave function describing the particle's
motion. The factor $\alpha$, homogeneous to a velocity, is not specified
by the theory, and might depend on the detector's characteristics.
We call the predicted distribution Eq.~(\ref{eq:density}) the \emph{density-like
flux}.
\item \emph{The quantum flux approach}: this approach originates from the
hydrodynamic interpretation~\citep{Madelung1927} of the Schrödinger
equation, where the probability density $\rho(\bm{r},t)=\vert\psi(\bm{r},t)\vert^{2}$
flows in the direction of the probability current, $\bm{j}=\frac{\hbar}{m}\text{Im}\left(\psi^{*}\bm{\nabla}\psi\right)$,
through the continuity equation $\dot{\rho}+\bm{\nabla}\cdot\bm{j}=0$.
According to this description, the distribution of times $t$ for
the particle to arrive onto some surface $S$ is the flux 
\begin{equation}
F(t)=\iint_{S}\bm{j}\cdot d\bm{S}\label{eq:probabilityFlux}
\end{equation}
of the probability current through the surface. As long as $F$ remains
positive, this description is consistent~\citep{McKinnon1995,Leavens1998}
with the Bohmian interpretation~\citep{Bohm1952,Bohm1952a} of quantum
mechanics (also known as de Broglie-Bohm's pilot wave theory), according
to which the particle has a definite trajectory that follows the probability
current. We call the predicted distribution Eq.~(\ref{eq:probabilityFlux})
the \emph{probability flux}.
\item \emph{The stochastic path approach}: in a different approach known
as Nelson's stochastic mechanics~\citep{Fenyes1952,Nelson1966,Nelson1985}
(which may be viewed as a variant of the pilot wave theory), the particle
follows a Brownian-motion-like trajectory whose arrival-time distribution
was shown~\citep{Naidon2024} to be approximately given by the flux
\begin{equation}
F_{-}(t)=\iint_{S}\bm{j}_{-}\cdot d\bm{S}\label{eq:backwardFlux}
\end{equation}
of the backward current $\bm{j}_{-}=\bm{j}-\bm{i}$, where $\bm{i}=\frac{\hbar}{m}\text{Re}\left(\psi^{*}\bm{\nabla}\psi\right)$.
We call the predicted distribution Eq.~(\ref{eq:backwardFlux}) the
\emph{backward flux}. Note that this result was obtained assuming
that the particle is detected on its first passage into the detector.
In the opposite limit where the particle can enter the detector many
times before being detected, the predicted arrival-time distribution
was shown~\citep{Nitta2008,Naidon2024} to be identical to the \emph{density-like
flux} Eq.~(\ref{eq:density}).
\item \emph{The axiomatic approach}: several works~\citep{Aharonov1961,Allcock1969b,Kijowski1974,Grot1996,Delgado1997,Anastopoulos2006}
arrived at yet another prediction, often called the \emph{Kijowski
distribution}~\citep{Kijowski1974,Muga2000}. For a planar surface
perpendicular to the $x$-axis and located at some coordinate $\Delta x$,
it reads~\citep{Leavens2005},
\begin{multline}
\qquad\Pi_{\text{K}}(t)\!=\!\frac{\hbar}{m}\sum_{\pm}\!\!\iint\!\frac{dk_{y}}{2\pi}\frac{dk_{z}}{2\pi}\times\\
\left|\int_{-\infty}^{\infty}\!\!\!\!\!dk_{x}\Theta(\pm k_{x})\left|k_{x}\right|^{\!\frac{1}{2}}e^{ik_{x}\Delta x}\tilde{\psi}(\bm{k},t)\right|^{2},\label{eq:KijowskiDistribution}
\end{multline}
where $\tilde{\psi}(\bm{k},t)\equiv\int d\bm{r}e^{-i\bm{k}\cdot\bm{r}}\psi(\bm{r},t)$
is the Fourier transform of $\psi$, and $\Theta$ is the Heaviside
step function. This distribution has been claimed to represent the
prediction of standard quantum theory~\citep{Aharonov1961,Allcock1969a,Allcock1969b,Allcock1969c,Kijowski1974,Werner1986,Grot1996,Delgado1997,Aharonov1998,Baute2000,Galapon2005,Juric2022},
although this has been disputed~\citep{Kijowski1999,Leavens2002,Egusquiza2003,Leavens2005}.
We will refer to it as the \emph{Kijowski flux}.
\end{enumerate}
We note that there is ongoing research~\citep{Allcock1969b,Aharonov1998,Muga1999,Damborenea2002,Navarro2003,Tumulka2022a,Tumulka2024,Goldstein2024b,Reddiger2026}
on whether or in which way the presence of the detector may affect
the above predictions.

\paragraph*{Gross-Pitaevskii description ---}

All the above predictions Eqs.~(\ref{eq:density}-\ref{eq:KijowskiDistribution})
can be calculated for any given one-body wave function $\psi$. In
our proposal we use a Bose- Einstein condensate. It is well known
that it can be described, to a very good approximation, by a single-particle
wave function $\psi$ obeying the Gross-Pitaevskii equation~\citep{Gross1961,Pitaevskii1961,Pethick2008},
\begin{equation}
i\hbar\dot{\psi}=\left(-\frac{\hbar^{2}\Delta}{2m}+V+\frac{4\pi\hbar^{2}a}{m}N\left|\psi\right|^{2}\right)\psi,\label{eq:Gross-Pitaevskii-equation}
\end{equation}
instead of the usual Schrödinger equation. Here, the interactions
between atoms are accounted for by the non-linear term proportional
to \textcolor{blue}{${\normalcolor a}$}, the s-wave scattering length,
which describes the strength of interactions between ultracold atoms.
The effects of these interactions on arrival times are investigated
in a companion study~\citep{Naidon2025b}, where it is shown that
they bring strong quantitative differences but no major qualitative
changes.

\paragraph*{Experimental setup ---}

\begin{figure}
\includegraphics[width=1\columnwidth]{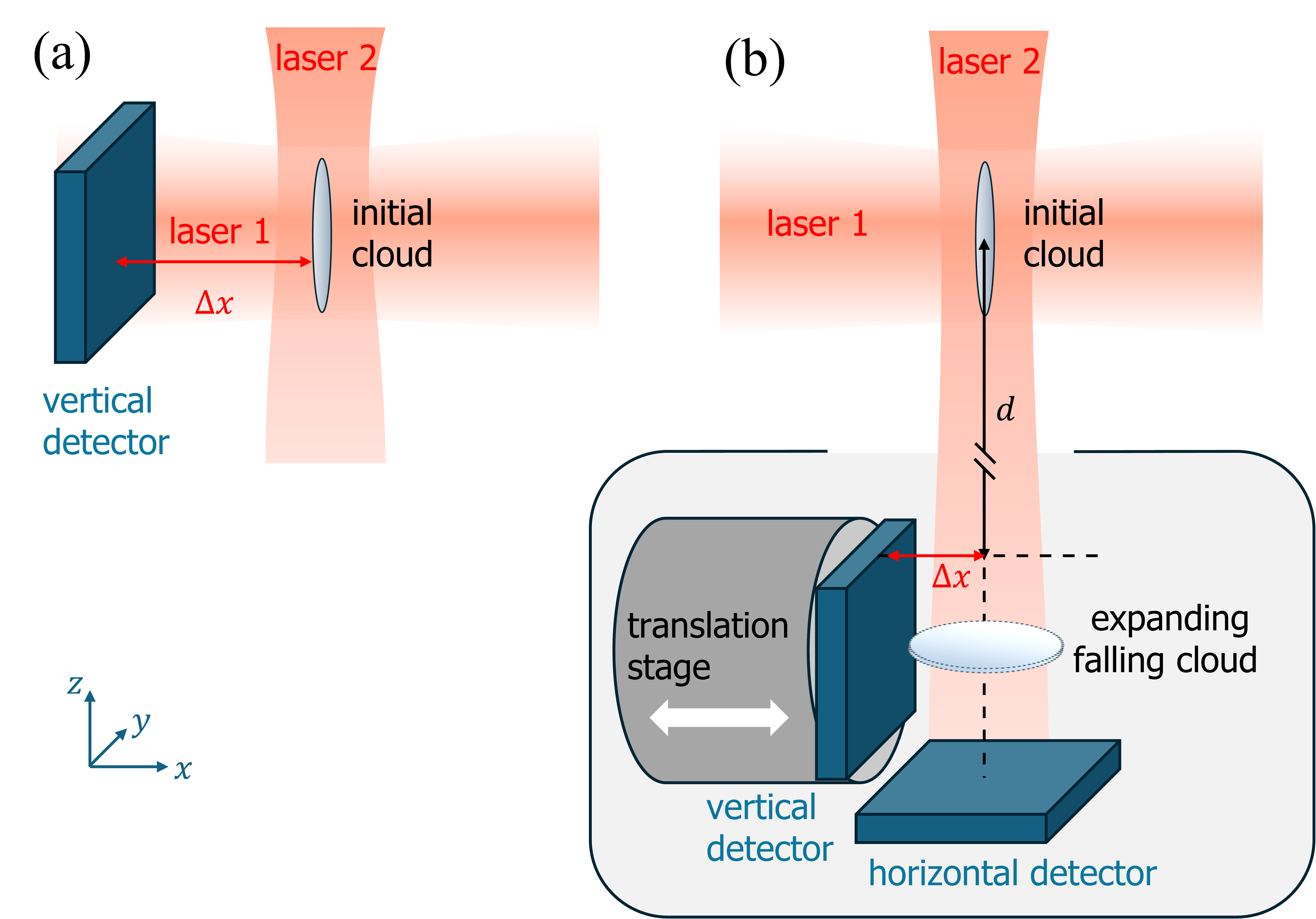}

\caption{\label{fig:setup}(a) Standard arrival-time experiment. (b) Our experimental
setup proposal. In both cases, a vertical and a horizontal laser beams
form an optical dipole trap in which a cloud of metastable helium-4
atoms is trapped -- here, it is illustrated for configuration (1)
of the lasers, which produces a vertical cigar-shaped cloud. At some
time, the cloud is released by switching off laser 2 (a) or both laser
beams (b). In our proposal (b), the cloud falls into a lower chamber
where it can be detected on a horizontal and a vertical microchannel
plate detector. }

\end{figure}

In quantum gas experiments, the standard detection schemes are fluorescence
and absorption imaging~\citep{Ott2016}, which are not suited for
arrival-time measurement at a fixed detector position. There exist
however two appropriate methods. Microchannel plate detectors have
been extensively used to detect metastable atoms~\citep{Vassen2012},
and thin sheets of light have also been used to detect the passage
of atoms~\citep{Buecker2009}. Our proposal is based on the first
method of detection. In such an experiment, illustrated in Fig.~\ref{fig:setup}a,
one prepares a condensate of metastable atoms in a trap (such as an
optical trap created by laser beams), switches off the trap to release
the atoms, and measures the time they take to hit a nearby microchannel
plate detector. However, such a setup has several issues for distinguishing
theories.

First, calculations show~\citep{Naidon2025b} that in order to observe
significant differences between the predictions of Eqs.~(\ref{eq:density}-\ref{eq:KijowskiDistribution}),
the detector should be very close to the trap, at distances of only
a few times the harmonic length of the trap. This poses not only
technical challenges to place a detector at such a precision close
to a trapped cloud, but also raises some fundamental concerns related
to the back action of the detector mentioned earlier\textcolor{blue}{.}

\begin{figure}[t]
\includegraphics[viewport=10bp 0bp 285bp 345bp,clip,width=1\columnwidth]{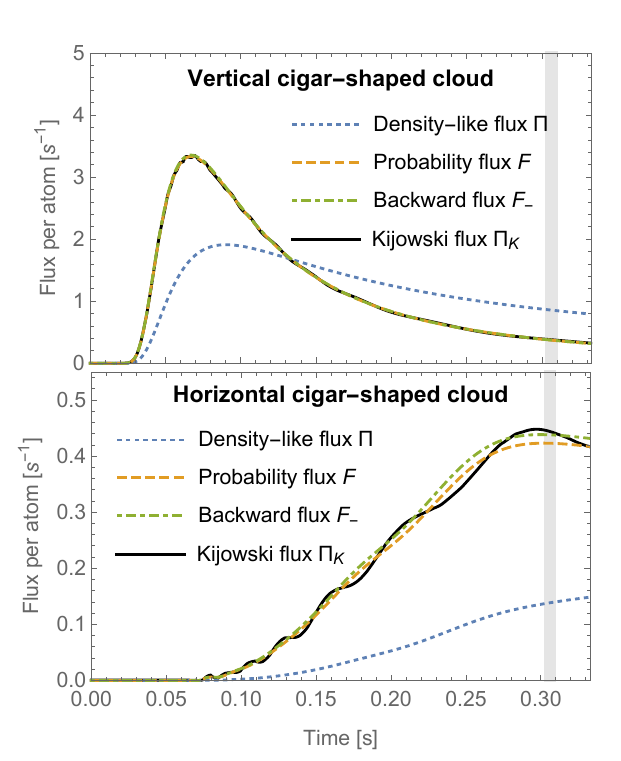}

\caption{\label{fig:TimeDistributions} Arrival-time distributions based on
the predictions Eqs.~(\ref{eq:density}-\ref{eq:KijowskiDistribution})
for a helium-4 atom in a condensate of $10^{4}$ atoms initially trapped
in a harmonic trap of transverse and axial frequencies $\omega_{\perp}=2\pi\times200\,\text{Hz}$
and $\omega_{\parallel}=2\pi\times25\,\text{Hz}$ in the vertical
(z) orientation (top panel) or horizontal (x) orientation (bottom
panel), and arriving onto an vertical detector located at a distance
$\Delta x=0.4\,\text{mm}$ from the trap -- see Fig.~\ref{fig:setup}a.
The coefficient $\alpha$ of Eq.~(\ref{eq:density}) is set to the
arbitrary values $\alpha=3.0\,\text{mm/s}$ (top) and $\alpha=0.4\,\text{mm/s}$
(bottom). In our proposed setup (Fig.~\ref{fig:setup}b), the detector
has a height $h=2\text{ cm}$, and is separated by a vertical distance
$d=45\,\text{cm}$ from the source of atoms. It can therefore probe
the above arrival-time distribution only in a small time window (shaded
area) corresponding to the cloud passing in front of the detector.}
\end{figure}

Second, the arrivals of the atoms cannot be observed in the vertical
direction, because gravity would strongly affect the arrival-time
distribution, making it indistinguishable from the semi-classical
falling flux $gt\iint_{S}dS\rho$~\citep{Naidon2025b}, where $S$
is the surface of the detector and $g$ is the Earth's gravity. 
Maintaining a vertical confinement of the cloud to avoid this issue,
and measuring instead its horizontal expansion on a vertical detector
could be a solution. However, the predicted signals only differ significantly
by their amplitude, not their shape. Figure~\ref{fig:TimeDistributions}
shows the predicted arrival-time distributions for a cigar-shaped
cloud of $N=10^{4}$ atoms in two orientations, (1) vertical (top
panel) and (2) horizontal (bottom panel), with axial trapping frequency
$\omega_{\parallel}=2\pi\times25\text{ Hz}$ and transverse trapping
frequency $\omega_{\perp}=2\pi\times200\text{ Hz}$, observed on an
idealised, vertically oriented planar detector of infinite size at
a horizontal distance of $\Delta x=0.4\text{ mm}$ from the trap centre.
In both cases, one can see that all predictions (except the Kijowski
flux $\Pi_{\text{K}}$ exhibiting small oscillations~\footnote{These oscillations are discussed in the companion study~\citep{Naidon2025b}.}
in the second case) are qualitatively similar. Since the initial state
of the system is never known precisely, extensive modelling and comparison
with the experimental data would be required to confidently exclude
theories.

To eliminate all the above issues, we propose a more practical and
effective setup depicted in Fig.~\ref{fig:setup}b. The first key
idea is to place the detector much further down from the trapped cloud,
at some vertical distance $d$. By switching off the lasers, the condensate
is released, falls, and hits the detector. Such a setup has already
been realised using a horizontal detector at $d\approx50$~cm~\citep{Vassen2012,Bouton2015}.
Here, however, we consider a vertical detector, so as to detect the
horizontal expansion of the cloud, which is not affected by gravity.
A horizontal detector can also be present for calibration purposes.

In this setup, the detectors are very far from the original location
of the cloud, which solves the first issues: not only is it more practical,
but it also induces minimal disturbance onto the system while it is
evolving; it is only at the very end of the expansion that the atoms
are detected. Additionally, it is not necessary to maintain a vertical
confinement in this setup.

In this situation, a finite-size detector will not record the full
arrival-time distribution of Fig.~\ref{fig:TimeDistributions} but
a tiny portion of it related to the detector size, the atomic cloud
size, and distance $d$. This time window is represented by a shaded
area in Fig.~\ref{fig:TimeDistributions}. The vertical detector
cannot detect horizontal arrivals outside that window. As can be seen,
it is too short for measuring any significant time variation of the
flux. 

To circumvent this limitation, we introduce the second key idea of
our setup: the vertical detector is placed onto a horizontal translation
stage enabling to change its position between experiments. This way,
the counts on the detector can be recorded for different horizontal
distances $\Delta x$ between the detector and the initial location
of the cloud. 

\begin{figure}[!t]
\includegraphics[viewport=12bp 0bp 316bp 394bp,clip,width=1\columnwidth]{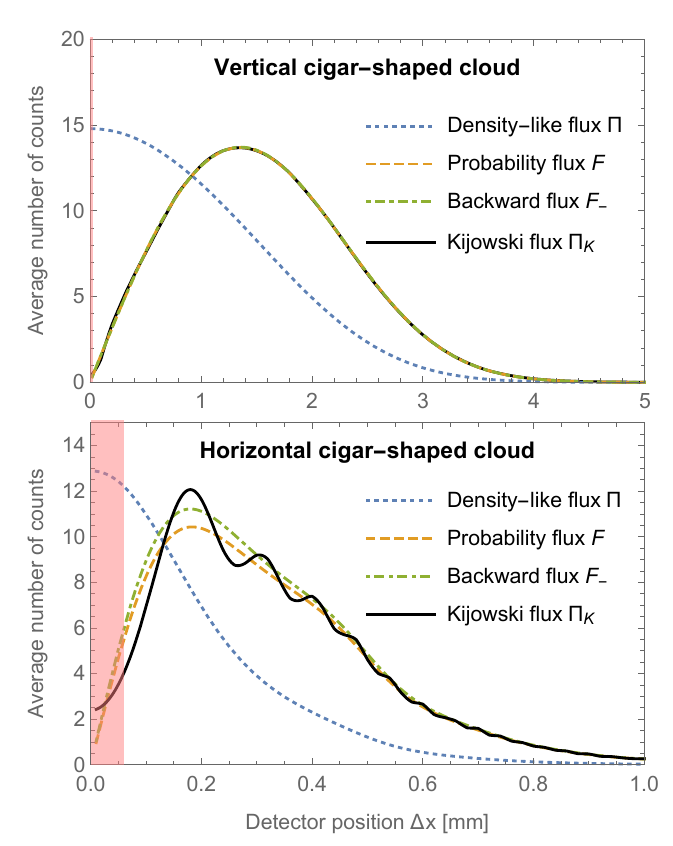}

\caption{\label{fig:Counts}Average number of counts on the vertical detector,
based on the predictions Eqs.~(\ref{eq:density}-\ref{eq:KijowskiDistribution}),
as a function of the horizontal position of the detector. The counts
at a given position correspond to a single experiment with $10^{4}$
atoms, taking into account a 25\% detection efficiency. The top panel
(bottom panel) shows the results for a cloud in the vertical (horizontal)
cigar configuration. Predictions are expected to become unreliable
for detector positions close to the initial radius of the cloud in
the $x$ direction, indicated by the red area on the left of each
plot, due to possible back action of the detector onto the cloud.}
\end{figure}

\paragraph*{Expected outcomes ---}

The predicted average total numbers of counts on the detector for
different detector positions are shown in Fig.~\ref{fig:Counts},
for a single experiment with $10^{4}$ atoms per detector position.
These numbers take into account the efficiency of the detector estimated
to be around 25\%. The numbers largely exceed the detector noise,
which is estimated to be less than $1\text{ count/cm}^{2}\text{/s}$,
i.e. 0.027 counts on average for a single experiment.

In the vertical cigar configuration (top panel of Fig.~\ref{fig:Counts}),
one can see that the density-like flux now exhibits a major qualitative
difference with respect to other predictions: its curve monotonically
decreases while the other curves exhibit a peak at some particular
position of the detector. This is not a coincidence due to the choice
of parameters, but a generic and stable feature resulting from the
fact that the density remains maximal at the origin, whereas the probability
and backward fluxes must vanish there. The prediction could thus easily
be confirmed or ruled out. This point showcases a key advantage of
our proposal to probe the arrival flux for different detector positions
rather than different times.

In the horizontal cigar configuration (bottom panel), one can observe
small oscillations in the Kijowski flux that are absent from other
predictions, similar to the oscillations visible in the bottom panel
of Fig.~\ref{fig:TimeDistributions}. This qualitative difference
could be used to confirm or rule out the axiomatic prediction. However,
the shot noise for a single experiment with $10^{4}$ atoms does
not permit to resolve these oscillations. One could prepare a condensate
with a larger number of atoms, although calculations show~\citep{supp}
that it may still be difficult to fully resolve the oscillations with
a single experiment per detector position. An efficient way to reduce
the statistical error is to repeat the experiment at the same detector
position. Despite the inevitable fluctuations of parameters between
experiments, such as the fluctuations in the initial number of atoms,
estimated to vary by up to 10\%, repeating the experiment about a
few hundred times for each detector position is expected to reduce
the statistical error to a degree where some oscillations become clearly
visible~\citep{supp}.

Although our proposed experiment could distinguish the density-like
flux (predicted by the quantum clock approach and stochastic trajectories'
multiple arrivals) and the Kijowski flux (axiomatic approach) from
other predictions, the probability flux (Bohmian trajectories' arrivals)
and backward flux (approximating stochastic trajectories' first arrivals)
may not be different enough to be distinguished from each other. This
may turn out to be irrelevant if the experimental outcomes rule out
these predictions altogether. Otherwise, a more advanced scheme will
be required, such as measuring arrival times of interfering clouds~\citep{Roncallo2023,AyatollahRafsanjani2023,Naidon2024}.
Note however that such experiments involving interferences enhance
discrepancies between the two theories due to the appearance of negative
flux. Such negative flux, which entails a possible reentry of particles
in the detector, make some predictions questionable (since negative
flux cannot be directly interpreted as a probability of arrival) and
more dependent on various assumptions about the detector and its effects
on the system. In contrast, our proposed experiment involves only
a single expanding cloud with only positive flux, which avoids all
these theoretical difficulties. We also stress that the experimental
signatures identified in Fig.~\ref{fig:Counts} occur at detector
positions relatively far from the initial location of the cloud.

In conclusion, we have proposed an experimental setup for indirectly
measuring the arrival times of a Bose-Einstein condensate of atoms,
which promises to unambiguously discriminate between at least three
major groups of predictions of arrival-time distribution. The proposal
is based on readily available technology and, unlike previous proposals,
minimises the unknown interactions between the detector and the system.\medskip{}

\begin{acknowledgments}P. N. acknowledges support from the JSPS Grants-in-Aid
for Scientific Research on Innovative Areas (No.JP23K03292). L. H.
is supported by the RIKEN special postdoctoral researcher program.
D. B. acknowledges founding from QuantERA Grant No. ANR-22-QUA2-000801
(MENTA), ANR Grant No. 20-CE-47-0001-01 (COSQUA), and the LabEx PALM
(No. ANR-10-LABX-0039PALM).\end{acknowledgments}

\bibliographystyle{apsrev4-2}
\bibliography{paper46}

\renewcommand{\theequation}{S\arabic{equation}} \renewcommand{\thefigure}{S\arabic{figure}} \renewcommand{\thetable}{S\arabic{table}} \onecolumngrid \setcounter{equation}{0} \setcounter{figure}{0} \setcounter{table}{0} \clearpage 

\section*{Supplemental material: ``Proposal of quantum arrival-time measurement
with a Bose-Einstein condensate''}

\begin{center}Pascal Naidon{ and }Lucas Happ

\selectlanguage{english}%
\emph{Few-Body Systems Physics Laboratory, RIKEN Nishina Centre, RIKEN,
Wakō, 351-0198 Japan}

\medskip{}

Denis Boiron

\emph{Université Paris-Saclay, Institut d’Optique Graduate School,
CNRS, Laboratoire Charles Fabry, 91127, Palaiseau, France}\end{center}

In an experiment, there are different kinds of fluctuations that can
affect the predictions presented in the main text, since these predictions
only indicate the expected average results. In this supplemental material,
we focus on the effect of theses fluctuations on the Kijowski flux
(predicted by the axiomatic approach) to highlight conditions under
which the oscillations, specific to that prediction, are observable.

Due to the limited number of atoms, there is a sampling error around
the predicted average values, known as shot noise. For a given average
count number $N_{i}$ predicted for the detector at position $i$,
the shot noise scales as $N_{i}^{1/2}$. In order to resolve the oscillatory
pattern of the axiomatic prediction, this shot noise should be smaller
than the amplitude of the oscillations.

\begin{figure}[h]
\includegraphics[width=8cm]{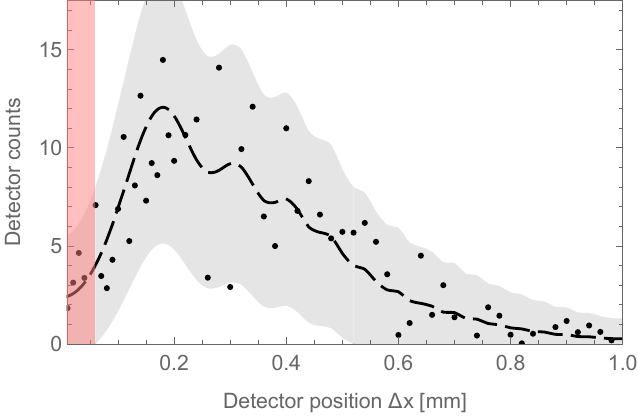}

\caption{\label{fig:shotNoise}Counts on the detector calculated using the
the axiomatic prediction for a condensate of $10^{4}$ atoms as considered
in the main text. The dashed curve shows the predicted average counts,
while the grey shaded area shows the amplitude of the corresponding
shot noise, and the dots represent a sample realization.}
\end{figure}

Figure.~\ref{fig:shotNoise} shows the expected range of shot noise
(grey shaded area) estimated by $\pm2N_{i}^{1/2}$ (2$\sigma$ deviation)
along with a sample realization, for the axiomatic prediction of the
bottom panel of Fig.~3 in the main text. It clearly shows that the
oscillations cannot be resolved in this case.

\begin{figure}[h]
\includegraphics[width=8cm]{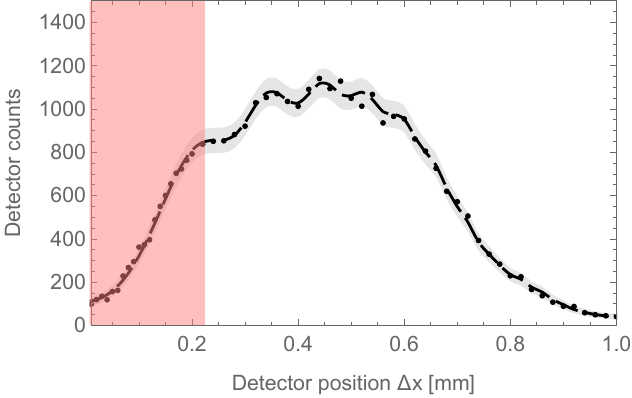}

\caption{\label{fig:shotNoise2}Same as Fig.~\ref{fig:shotNoise} for a condensate
with $N=10^{6}$ atoms, initially held in a trap with frequencies
$\omega_{\perp}=2\pi\times200\,\text{Hz}$ and $\omega_{\parallel}=2\pi\times15\,\text{Hz}$.}
\end{figure}

Here, the total number of atoms is $10^{4}$. Increasing the total
number of atoms in the condensate can help reduce the shot noise,
although this affects interactions between the atoms, resulting in
a smaller contrast of the oscillatory pattern. In Fig.~\ref{fig:shotNoise2},
the number of atoms in the condensate is set to $10^{6}$ and the
axial frequency is changed accordingly to $\omega_{\parallel}=2\pi\times15\,\text{Hz}$
in order to obtain the best contrast of the oscillations. The oscillatory
pattern now becomes apparent. However, it remains noisy and stands
at distances close to the initial size of the cloud shown by the red
area in Fig.~\ref{fig:shotNoise2}.

The best way to reduce the shot noise is to repeat the experiment
many times at each detector position. In this case, we keep the number
of atoms at about $10^{4}$ to maintain a higher contrast of the oscillatory
pattern, and average the results over many experiments. Ideally, the
experiments should be performed in exactly the same conditions. However,
due to instabilities in the lasers, there are always fluctuations
in the initial number of atoms, estimated to be of the order of 10\%,
as well as fluctuations of trapping frequencies, estimated to be of
the order of 2\%.

\begin{figure}
\includegraphics[width=8cm]{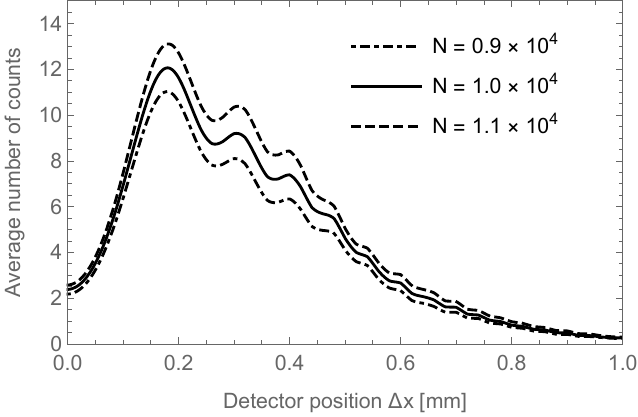}\caption{\label{fig:variation-with-N}Average number of counts on the detector
calculated with the axiomatic prediction for three different values
of the total number of atoms $N$.}
\end{figure}

\begin{figure*}
\includegraphics[width=8cm]{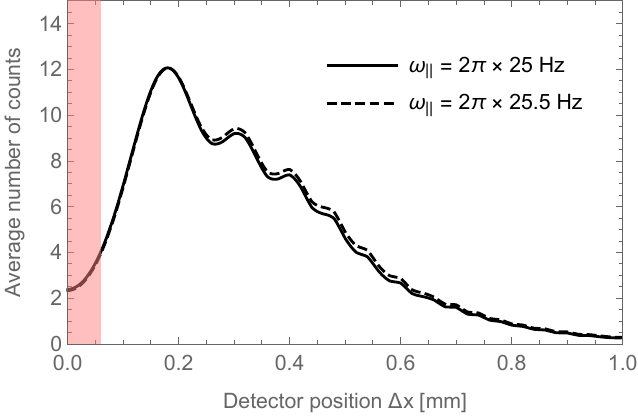} \includegraphics[width=8cm]{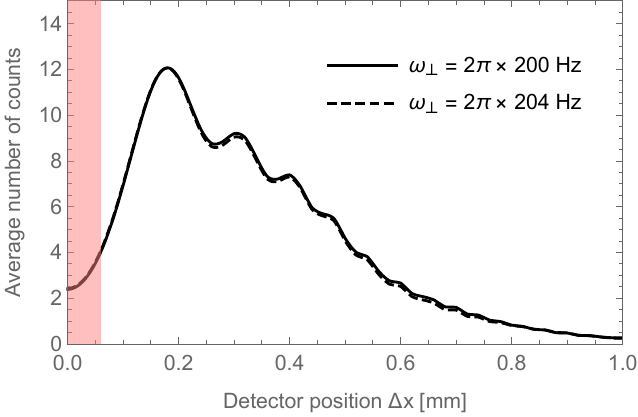}\caption{\label{fig:variation-with-omega-parallel}Average number of counts
on the detector calculated with the axiomatic prediction for two different
values of the trapping frequencies $\omega_{\parallel}$ (left) and
$\omega_{\perp}$ (right).}
\end{figure*}

To visualise the effect of these fluctuations, we plot in Fig.~\ref{fig:variation-with-N}
the average number of counts calculated with the axiomatic prediction
for three different values of the total number of atoms, spanning
the expected fluctuation range. One can see that the oscillations
remain mostly in phase over this range. We also plot in Fig.~\ref{fig:variation-with-omega-parallel}
the variations with the trapping frequencies. One can see that these
variations make little change to the average number of counts.

\begin{figure*}
\includegraphics[width=8cm]{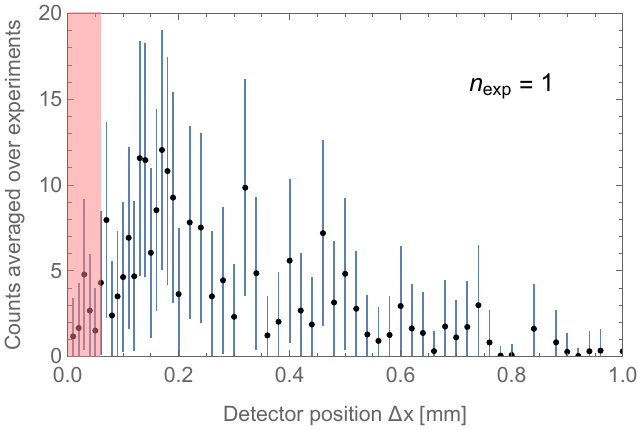}\includegraphics[width=8cm]{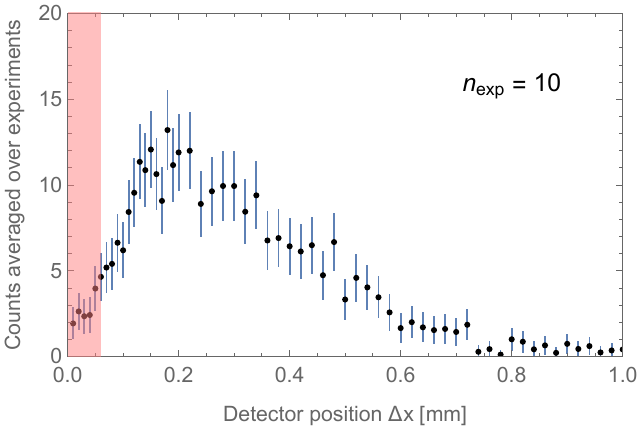}

\includegraphics[width=8cm]{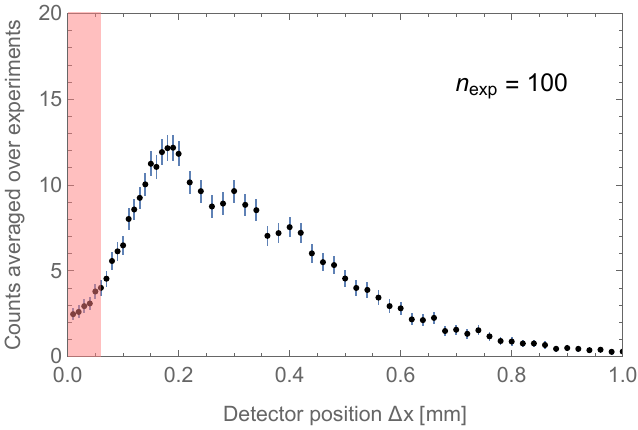}\includegraphics[width=8cm]{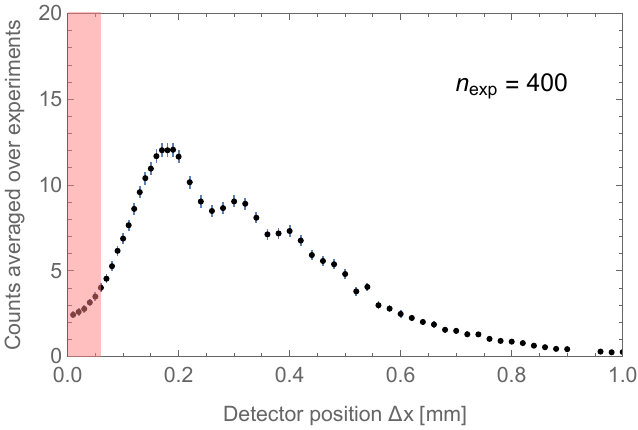}

\caption{\label{fig:experiments}Simulations of the number of counts on the
detector averaged over $n_{\text{exp}}$ experiments, each panel corresponding
to a different value of $n_{\text{exp}}$. These simulations take
into account fluctuations of the experimental parameters between experiments,
based on the results of Figs.~\ref{fig:variation-with-N} and \ref{fig:variation-with-omega-parallel}.
The error bars show the 2$\sigma$ (68\%) confidence interval $\pm2\sqrt{N_{i}^{\text{tot}}}/n_{\text{exp}}$
one could estimate from the data, with $N_{i}^{\text{tot}}$ being
the cumulated number of counts on the detector for all experiments
at detector position $i$. }

\end{figure*}

Finally, we simulate the outcome of averaging many experiments from
these results. One can check that over the range of variation of the
parameters $(N,\omega_{\parallel},\omega_{\perp})$ the results vary
almost linearly so that they can be interpolated linearly. We can
then simulate fluctuations of the parameters $(N,\omega_{\parallel},\omega_{\perp})$
by random variables following a normal distribution with standard
deviations (10\%,~2\%,~2\%). For each set of parameters thus obtained,
corresponding to a single experiment, we obtain the predicted average
number of counts by linear interpolation of our previous results.
Then, we simulate the shot noise of that experiment by adding to each
average count a normally distributed random fluctuation with standard
deviation equal to the square root of the average count. Repeating
this procedure to simulate many experiments, we obtain averaged results
shown in Fig.~\ref{fig:experiments} for different numbers of experiments.
We conclude that averaging over a few hundreds of experiments would
reveal the oscillatory pattern of the axiomatic prediction.

 \end{document}